\documentclass[10pt,aps,prb,twocolumn,showpacs,preprintnumbers,amsmath,amssymb,superscriptaddress,floatfix]{revtex4-1}

\usepackage[normalem]{ulem}
\usepackage{graphicx}
\usepackage{dcolumn}
\usepackage{amsmath}
\usepackage{color}
\usepackage{multirow}
\usepackage[hidelinks]{hyperref}
\usepackage[range-units=single, range-phrase={--}]{siunitx}
\usepackage{csquotes}

\newcommand{\msr}{$\mu$SR}
\newcommand{\lem}{LE-\msr}

\newcommand{\smb}{SmB$_6$}
\newcommand{\etal}{\emph{et al.}}
\newcommand{\usim}{\sim \!}
\newcommand{\e}{\mathrm{e}}

\newcommand{\mysection}[1]{\smallskip\emph{#1 ---}\phantomsection}

\bibpunct{[}{]}{,}{n}{}{}

\begin{document}
\title{Suppression of magnetic excitations near the surface of the
  topological Kondo insulator \texorpdfstring{\smb{}}{SmB6}}
\author{P.~K.~Biswas}
\email{pabitra.biswas@stfc.ac.uk}
\affiliation{Laboratory for Muon Spin Spectroscopy, Paul Scherrer Institut, CH-5232 Villigen PSI, Switzerland}
\affiliation{ISIS Pulsed Neutron and Muon Source, STFC Rutherford
  Appleton Laboratory, Harwell Campus, Didcot, Oxfordshire, OX11 0QX,
  United Kingdom}
\author{M. Legner}
\affiliation{Institut f\"ur Theoretische Physik, ETH Z\"urich, 8093 Z\"urich, Switzerland}
\author{G.~Balakrishnan}
\affiliation{Department of Physics, University of Warwick, Coventry, CV4 7AL, UK}
\author{M.~Ciomaga~Hatnean}
\affiliation{Department of Physics, University of Warwick, Coventry, CV4 7AL, UK}
\author{M.~R.~Lees}
\affiliation{Department of Physics, University of Warwick, Coventry, CV4 7AL, UK}
\author{D.~McK.~Paul}
\affiliation{Department of Physics, University of Warwick, Coventry, CV4 7AL, UK}
\author{E.~Pomjakushina}
\affiliation{Laboratory for Scientific Developments and Novel Materials, Paul Scherrer Institut, CH-5232 Villigen PSI, Switzerland}
\author{T.~Prokscha}
\author{A.~Suter}
\affiliation{Laboratory for Muon Spin Spectroscopy, Paul Scherrer Institut, CH-5232 Villigen PSI, Switzerland}
\author{T. Neupert}
\affiliation{Physik-Institut, Universit\"at Z\"urich, Winterthurerstrasse 190, 8057 Z\"urich, Switzerland}
\author{Z.~Salman}
\email{zaher.salman@psi.ch}
\affiliation{Laboratory for Muon Spin Spectroscopy, Paul Scherrer Institut, CH-5232 Villigen PSI, Switzerland}

\begin{abstract}
  We present a detailed investigation of the temperature and depth
  dependence of the magnetic properties of 3D topological Kondo
  insulator \smb{}, in particular near its surface. We find that local
  magnetic field fluctuations detected in the bulk are suppressed
  rapidly with decreasing depths, disappearing almost completely at
  the surface. We attribute the magnetic excitations to spin excitons
  in bulk \smb{}, which produce local magnetic fields of about $\usim
  \SI{1.8}{mT}$ fluctuating on a time scale of $\usim
  \SI{60}{\ns}$. We find that the excitonic fluctuations are
  suppressed when approaching the surface on a length scale of
  \SIrange{40}{90}{\nm}, accompanied by a small enhancement in static
  magnetic fields. We associate this length scale to the size of the
  excitonic state.
\end{abstract}

\maketitle

\mysection{Introduction}
Topological Insulators (TIs) are a class of quantum materials that are
characterized by a fully insulating gap in the bulk and robust
metallic topological surface states. It was suggested that these are
promising materials for electronic spin
manipulation~\cite{Moore2010n}. Theoretical studies predicted that the
prototypical Kondo insulator \smb{} belongs to this new class of
materials~\cite{Dzero2010prl,Alexandrov2013prl,Lu2013prl,Dzero2012prb}. This
was later supported by
transport~\cite{Wolgast2013prb,Kim2013sr,Zhang2013prx,Kim2014nm} and
angle-resolved photoemission spectroscopy
(ARPES)~\cite{Xu2013prb,Neupane2013nc,Jiang2013nc} measurements.  Xu
\etal{} have also revealed that the surface states of \smb{} are spin
polarized~\cite{Xu2014nc}, where the spin is locked to the crystal
momentum, respecting time reversal and crystal symmetries. At high
temperatures, Kondo insulators behave as highly correlated metals,
while at low temperature they are insulators due to the formation of
an energy gap at the Fermi
level~\cite{Aeppli1992ccmp,Riseborough2000adp,Coleman2007}.  The
opening of a gap at low temperature is attributed to the hybridization
between the localized $f$-electrons and the conduction
$d$-electrons. In \smb{}, the resistivity increases exponentially as
the temperature is decreased, as expected for a normal
insulator. However, as the temperature is decreased below \SI{4}{\K},
the resistivity saturates at a finite value (a few \SI{}{\ohm
  \cm})~\cite{Menth1969prl,Allen1979PRB}. This behavior was attributed
to extended states~\cite{Cooley1995prl}, whose nature was
revealed recently by transport experiments, identifying them with
metallic surface
states~\cite{Wolgast2013prb,Kim2013sr,Zhang2013prx,Kim2014nm} and
supporting predictions of the nontrivial topological nature of \smb.
ARPES measurements reveal a Kondo gap of $\usim\SI{20}{\meV}$ in the
bulk and identify the low-lying bulk in-gap states close to the Fermi
level~\cite{Xu2013prb,Neupane2013nc,Jiang2013nc,Miyazaki2012prb,Denlinger2000pb}. These
in-gap states have been associated with magnetic
excitations~\cite{Alekseev1993pb,Riseborough2000adp,Miyazaki2012prb}
and found to disappear as the temperature is raised above
\SIrange{20}{30}{\K}~\cite{Alekseev1993pb,Neupane2013nc}. Other ARPES
results suggest that the transition is very broad and that the in-gap
states disappear completely at a much higher
temperature~\cite{Xu2013prb}, or that they gradually transform from 2D
to 3D nature with increasing
temperature~\cite{Denlinger2014jpscs,Denlinger2013arxiv}.

The magnetic properties of bulk \smb{} have been extensively studied
using magnetization measurements~\cite{Menth1969prl}, inelastic
neutron
scattering~\cite{Alekseev1993pb,Alekseev1995jpcm,Alekseev2010jetp,Fuhrman2015prl},
nuclear magnetic resonance
(NMR)~\cite{Takigawa1981jpsj,Caldwell2004,Caldwell2007prb,Schlottmann2014prb}
and muon spin relaxation (\msr)~\cite{Biswas2014prb}. These
measurements detected magnetic excitations at energies below the bulk
gap. However, magnetic ordering in the bulk of \smb{} was ruled out by
magnetization~\cite{Menth1969prl} and \msr\ measurements down to
\SI{20}{mK}~\cite{Biswas2014prb} (except under high
pressure~\cite{Barla2005prl}). In contrast, low temperature
magnetotransport measurements indicate magnetic ordering at the
surface of \smb{} below $\SI{600}{mK}$, which was attributed to
ferromagnetic~\cite{Nakajima2016np} or possibly
glassy~\cite{Wolgast2015prb} ordering. This ordering is claimed to
involve Sm\textsuperscript{3+} magnetic moments which were detected
using x-ray absorption spectroscopy (XAS) at the surface of
\smb{}~\cite{Phelan2014prx}.

Although various theoretical
\cite{Dzero2010prl,Alexandrov2013prl,Lu2013prl,Dzero2012prb} and
experimental
\cite{Wolgast2013prb,Kim2013sr,Zhang2013prx,Kim2014nm,Xu2013prb,Neupane2013nc,Jiang2013nc,Xu2014nc}
studies have now established compelling evidence that \smb{} is a
topological Kondo insulator, a number of open questions remain
unanswered. In particular, the source of the magnetic excitations
mentioned above is still unclear. It was suggested that an
\emph{excitonic state} is responsible for these
fluctuations~\cite{Riseborough2000adp, Alekseev2010jetp,
  Fuhrman2015prl,Knolle2016arxiv}. In this context, it is also
important to understand the interplay between these magnetic
excitations and the topological surface sates in order to elucidate
the source of reported magnetic ordering at the surface of
\smb{}~\cite{Nakajima2016np,Wolgast2015prb}. In this paper, we address
these important aspects using depth-resolved low-energy \msr{} (\lem)
measurements on single-crystal samples of \smb{}. We detect a clear
signature of fluctuating local magnetic fields appearing below
$\usim\SI{15}{\K}$, similar to our previous bulk
measurements~\cite{Biswas2014prb}. The typical size of the fluctuating
field in the bulk is $\usim \SI{1.8}{mT}$ with a correlation time of
$\usim \SI{60}{\ns}$. Moreover, we find that the magnitude and/or
fluctuation time of these magnetic fields decreases gradually near the
surface, over a length scale of \SIrange{40}{90}{\nm}, possibly
disappearing completely at the surface of \smb{}. We propose that
excitonic states are responsible for these fluctuating magnetic
fields. In contrast, we detect an enhancement of static magnetic
fields near the surface, which may be attributed to an increasing
number of Sm\textsuperscript{3+} moments at the surface of \smb{}
\cite{Phelan2014prx}.

\mysection{Experimental details}\label{sec:experimental}
\msr\ measurements were performed using the
LEM~\cite{Morenzoni1994prl,Prokscha2008nima} and DOLLY spectrometers
at PSI, Switzerland. In these measurements, \SI{100}{\%} spin
polarized positive muons are implanted into the sample. The evolution
of the spin polarization, which depends on the local magnetic fields,
is monitored via the anisotropic beta decay positron which is emitted
preferentially in the direction of the muon's spin at the time of
decay. Using appropriately positioned detectors one can measure the
asymmetry, $A(t)$, of the beta decay along the initial polarization
direction. $A(t)$ is proportional to the time evolution of the spin
polarization of the ensemble of implanted spin
probes~\cite{Yaouanc2010}. Conventional \msr\ experiments use surface
muons with implantation energy of \SI{4.1}{MeV}, resulting in a
stopping range in typical density solids from \SI{0.1}{mm} to
\SI{1}{mm}. Thus limiting their application to studies of bulk
properties, i.e., they cannot provide depth-resolved information or
study extremely thin film samples. Depth-resolved \msr{} measurement
can be performed at the LEM spectrometer using muons with tunable
energies in the \SIrange{1}{30}{\keV} range, corresponding to
implantation depths of \SIrange{10}{200}{\nm}. All the \msr{} data
reported here were analyzed using the MUSRFIT
package~\cite{Suter2012pp}.

The studied single crystals of \smb{} samples were grown using the
floating-zone method~\cite{CiomagaHatnean2013sr}. \lem{} measurements
were performed on a mosaic of 6 disc shaped single crystals, aligned
with their [100] axis normal to the surface, and glued on a silver
backing plate. The bulk \msr\ measurements reported here were
performed on one of these single crystals.

\begin{figure}[hb]
\includegraphics[width=0.9\linewidth]{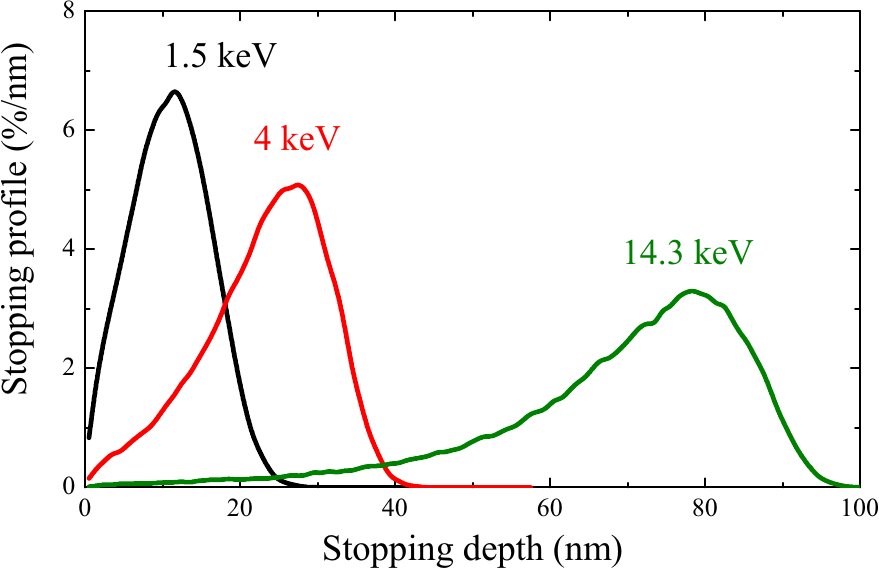}
\caption{(Color online) Muon implantation profiles in \smb{},
  calculated using TRIM.SP for various implantation energies.}
 \label{fig:stopping_profile}
\end{figure}

\mysection{Results}\label{sec:results}
\begin{figure}[hbt]
\includegraphics[width=0.8\linewidth]{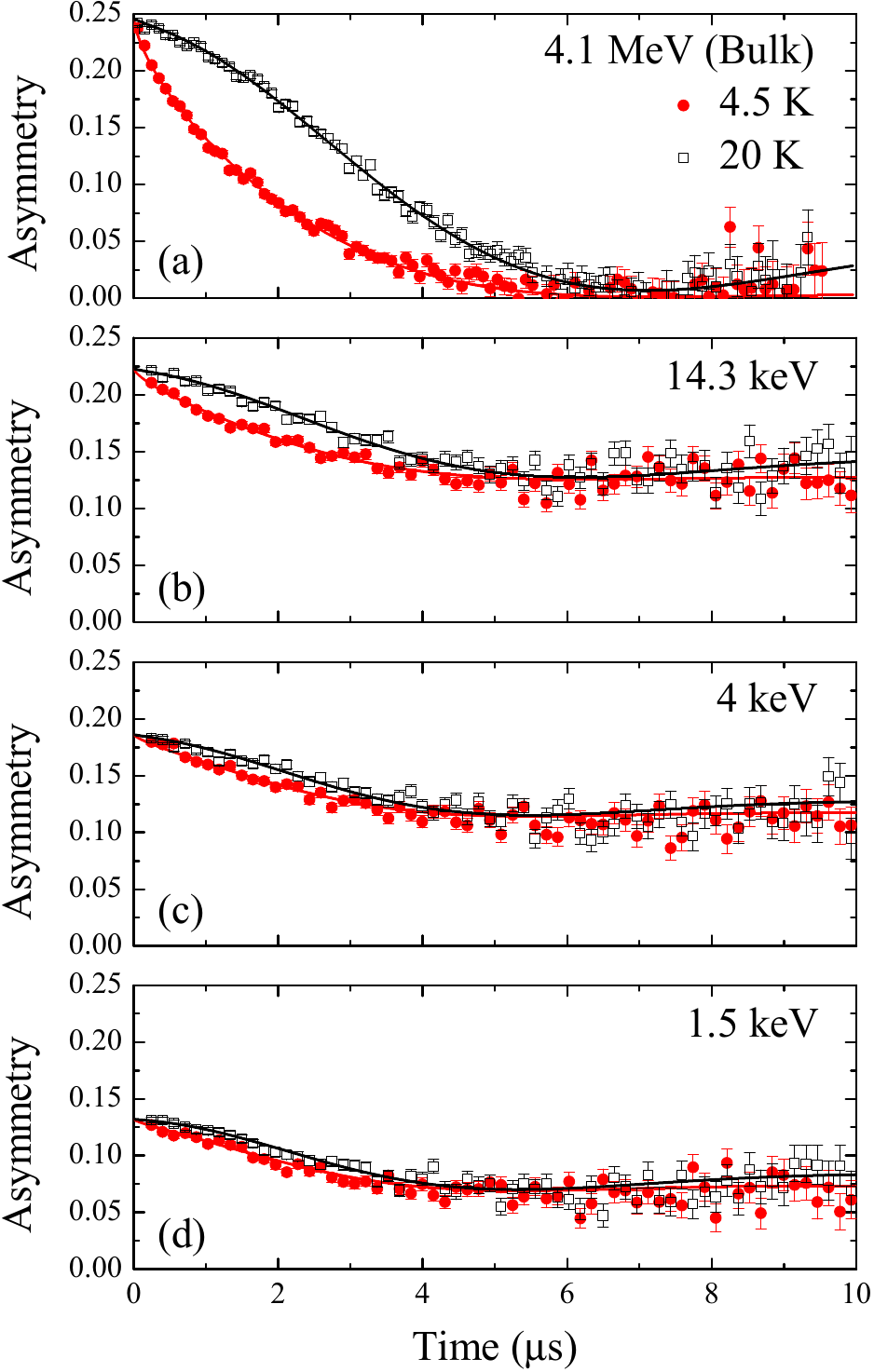}
\caption{(Color online) ZF-\msr{} spectra obtained at different
  temperatures and implantation energies. The solid lines are fits to
  Eq.~(\ref{eq:KT_ZFequation}).}
 \label{fig:asymmetry}
\end{figure}
Figure~\ref{fig:asymmetry}(a) shows typical zero field (ZF) \msr\
asymmetries, measured at two different temperatures, above and below
the \enquote{critical} temperature $\usim\SI{15}{K}$, i.e., where
strong local magnetic field fluctuations appear in bulk
\smb{}~\cite{Biswas2014prb}. These are compared in
Fig.~\ref{fig:asymmetry}(b--d) to \lem\ measurements at the same
temperatures and three different muon implantation energies, $E$. The
corresponding muon stopping profiles in \smb{} for the different
energies, which were calculated using a Monte Carlo program
TRIM.SP~\cite{Morenzoni2002nimb}, are shown in
Fig.~\ref{fig:stopping_profile}.
At \SI{20}{K} we observe a Gaussian-like muon spin damping for all
four different energies. This type of damping is attributed to
randomly oriented static magnetic fields~\cite{Yaouanc2010}, which
reflect the Gaussian field distribution typically produced by dipolar
fields from nuclear moments (static on the time scale of \msr). A
clear change in the shape of the asymmetry is detected upon cooling
(from Gaussian to Lorentzian), which indicates the appearance of
additional dilute local magnetic fields and a change in the internal
field distribution \cite{Uemura1985prb}. Since the dipolar fields from
nuclear moments do not change with temperature, we argue that the
appearance of additional dilute local magnetic fields is most probably
due to electronic magnetic moments in \smb{} which are dynamic in
nature within the \msr\ time scale \cite{Biswas2014prb}. Most
importantly, however, we find that the difference between the low and
high temperature asymmetries becomes less pronounced with decreasing
$E$, i.e., as we approach the surface of the \smb{}. As we discuss
below, this indicates that the size and/or fluctuation time of the
observed magnetic fields at low temperatures decreases gradually with
decreasing depth.

\begin{figure}[htb]
\includegraphics[width=0.9\linewidth]{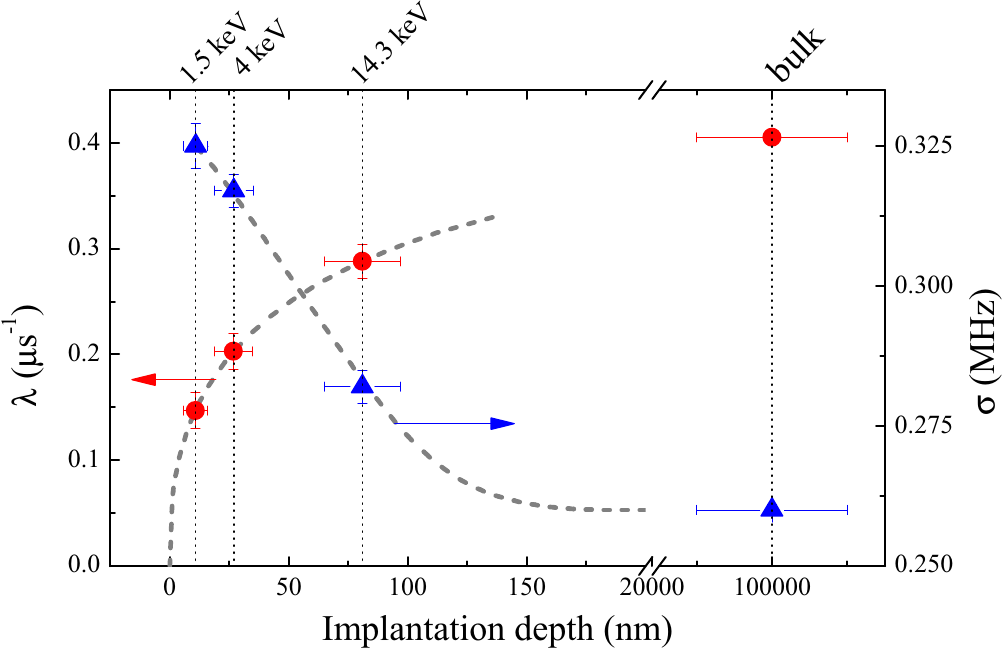}
\caption{(Color online) The relaxation rates $\lambda$ at
  $\usim\SI{4.5}{K}$ (red, left axis) and $\sigma$ (blue, right axis)
  as a function of muon implantation depth in \smb{}. The dashed lines
  are guides to the eye and the dotted vertical lines indicate the
  different $E$ values.}
 \label{fig:lambda_implantation}
\end{figure}

We turn now to a quantitative analysis of our \msr\ data. Following
the same analysis procedure used previously for the bulk
measurements~\cite{Biswas2014prb}, all ZF spectra can be fitted well
using a Gaussian Kubo-Toyabe relaxation function multiplied by a
stretched exponential decay function,
\begin{equation}
  A(t)=
  A_0\left\{\frac{1}{3}+\frac{2}{3}\left(1-\sigma^2t^2\right)\e^{-\frac{\sigma^2t^2}{2}}\right\}\e^{-(\lambda
  t)^\beta} +A_{\rm bg},
 \label{eq:KT_ZFequation}
\end{equation}
where $A_0$ is the initial asymmetry, $\beta$ is the stretch
parameter, and $A_{\rm bg}$ is a non-relaxing background
contribution. $\sigma$ is the width of static field distribution,
e.g., due to nuclear moments, while $\lambda$ is the muon spin
relaxation rates due to the presence of dynamic local fields. Note,
$A_{\rm bg}=0$ in the bulk \msr\ measurements, but it is non-zero in
the \lem\ measurements due to muons missing the sample and landing in
the silver backing plate.
\begin{figure}[htb]
\includegraphics[width=0.9\linewidth]{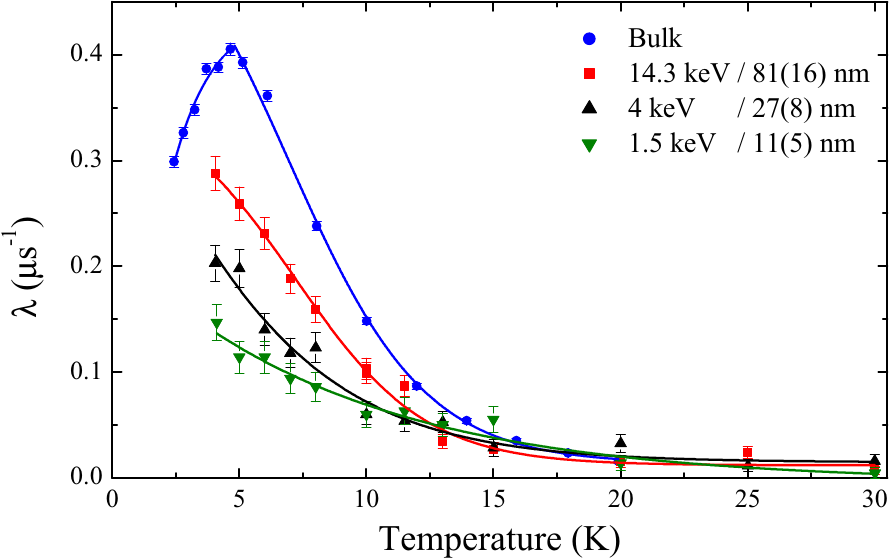}
\caption{(Color online) Temperature dependence of the dynamic muon
  spin relaxation rate $\lambda$ for different muon implantation
  energies. The solid lines are guides to the eye.}
 \label{fig:lambda_temp}
\end{figure}
For consistency with the bulk-\msr{} data
analysis~\cite{Biswas2014prb}, we maintain $A_0$, $\sigma$ and $A_{\rm
  bg}$ as globally common variables for all temperatures at a
particular muon implantation energy. Similarly, we also keep the value
obtained from bulk measurements, $\beta=0.72(1)$, fixed for all
temperatures and implantation energies. Figure~\ref{fig:lambda_temp}
shows the obtained $\lambda$ values from the fit as a function of
temperature for each implantation energy/depth. We observe a large
increase in $\lambda$ below $\usim\SI{15}{K}$ in the bulk-\msr\ data
with a pronounced peak at $\usim\SI{4.5}{K}$ which we attribute to
gradual slowing down in the dynamics of the local magnetic fields at
low temperatures~\cite{Biswas2014prb}. As expected from our
qualitative discussion above, we observe similar increase in $\lambda$
below $\usim\SI{15}{K}$ for all other implantation energies, though it
becomes less pronounced as we approach the surface. Although a gradual
slowing down is observed for all implantation energies, our results
clearly show that the nature of magnetic fluctuations strongly depends
on depth. In Fig.~\ref{fig:lambda_implantation} we plot $\lambda$ at
$\usim\SI{4.5}{K}$ and $\sigma$ as a function of the muon implantation
depth in \smb{}. The relaxation rate $\lambda$ decreases rapidly with
decreasing depth and may be extrapolated to $\lambda \rightarrow 0$ at
the surface of \smb{} (dashed line).
This is accompanied by an increase in $\sigma$ with decreasing
depth. The value of $\sigma$ in the bulk is consistent with what we
expect from (predominantly boron) nuclear magnetic moments in this
system \cite{Biswas2014prb}. Therefore, the observed increase near the
surface must be due to additional sources of relatively small and
\emph{static} magnetic fields. This may be due to an increased
concentration of Sm$^{3+}$ moments near the surface of \smb{} which
was observed in XAS measurements~\cite{Phelan2014prx}. The increase in
$\sigma$ may hint to a possible magnetic ordering at the surface of
\smb{}, such as that reported below
$\SI{600}{mK}$~\cite{Nakajima2016np,Wolgast2015prb}. However, this
cannot be fully confirmed since it is not possible to reach the
required low temperatures in \lem{} measurements.

Note that $\lambda$ reflects the spin lattice relaxation rate of the
muon spin in ZF, which is proportional to $\Delta B^2 \tau$, where
$\Delta B$ is the size of the fluctuating local field sensed by the
implanted muons and $\tau$ is its correlation time. Therefore, the
observed decrease in $\lambda$ at lower implantation energies may be
attributed to a decrease in $\Delta B$ and/or $\tau$ as we approach
the surface of \smb{}.
\begin{figure}[htb]
\includegraphics[width=0.9\linewidth]{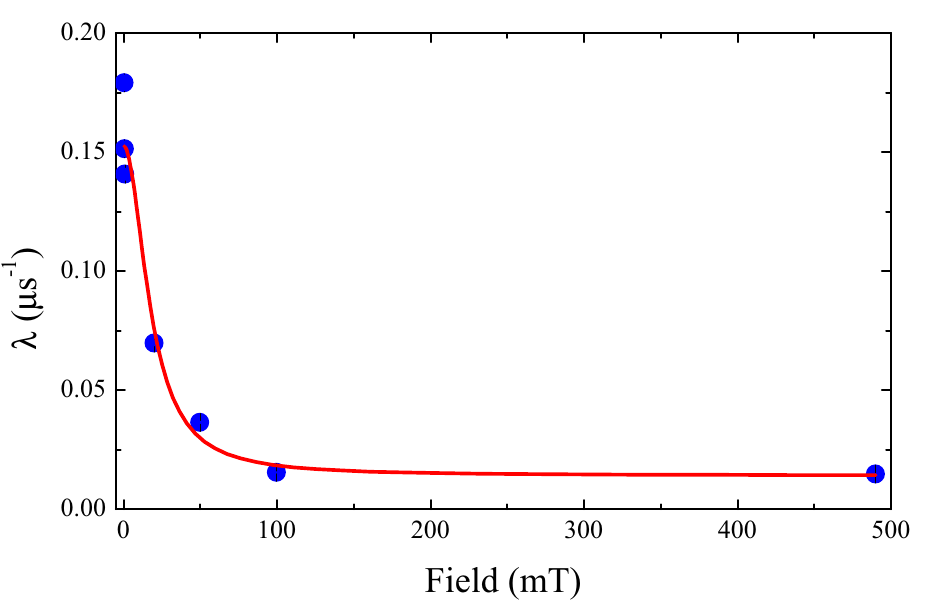}
\caption{(Color online) The relaxation rate $\lambda$ at \SI{1.8}{K}
  as a function of applied field~\cite{Biswas2014prb}. The solid line
  is the fit described in the text.}
  \label{fig:lambda_field}
\end{figure}
To evaluate the size of $\Delta B$ and $\tau$ we measure the asymmetry
as a function of longitudinal magnetic field, i.e., applied along the
direction of initial muon spin polarization. The field dependence of
$\lambda$ follows~\cite{Uemura1985prb,Keren1994prb,Salman2002prb},
\begin{equation} \label{lamvsB}
\lambda=\frac{2 \tau (\gamma \Delta B)^2}{1+(\tau \gamma B_0)^2},
\end{equation}
where $\gamma=2 \pi \times \SI{135.5}{MHz\per T}$ is the gyromagnetic
ratio of the muon and $B_0$ is the applied magnetic field. The
experimental results, which were measure in the bulk of \smb{} at
\SI{1.8}{K}, fit well to Eq.~\eqref{lamvsB} (see
Fig.~\ref{fig:lambda_field}) giving $\Delta B =\SI{1.8(2)}{mT}$ and
$\tau=\SI{60(10)}{ns}$~\cite{LCRnote}. Here we assume that we are in
the fast fluctuations limit, $\tau \gamma \Delta B \ll 1$, which is
consistent with the values obtained from the fit.

\mysection{Discussion}\label{sec:discussion}
We now discuss our data assuming bulk excitons as a source for the
observed magnetic fluctuations.  The excitons are believed to be of
antiferromagnetic nature with a wavelength of the order of a few
lattice constants~\cite{Riseborough2000adp, Fuhrman2015prl}. The
observed decay length of magnetic fluctuations near the surface
(\SIrange{40}{90}{nm}) should then be interpreted as the coherence
length or \enquote{size} of the excitons.  This size is much larger
than the ordering wavelength so that the exciton can be thought of as
a fluctuating region with antiferromagnetic correlations.

The measured value $\Delta B\sim \SI{1.8}{mT}$ for the width of the
distribution of magnetic fields can be used to estimate the magnitude
of the fluctuating magnetic moments. We assume that the muons stop at
a random position inside the cubic unit cell of \smb{} with a lattice
constant of \SI{4.13}{\angstrom}. Calculating the distribution of
magnetic fields due to the antiferromagnetic correlations in the
region of the exciton yields an average value of $\usim 0.01
\mu_\mathrm{B}$ for the magnetic moments, where $\mu_\mathrm{B}$ is
the Bohr magneton.

In addition, we can use a simple hydrogen model for the exciton,
describing it as a bound state of an electron and a hole, in order to
relate the size to the reduced mass $\mu_\mathrm{ex}$ of the
electron--hole pair via $d=a_0
\epsilon_{\mathrm{r}}m_\mathrm{e}/\mu_\mathrm{ex}$. Here, $a_0$ is the
Bohr radius and $\epsilon_\mathrm{r}$ is the dielectric constant of
\smb{}, which is estimated between $\epsilon_{\mathrm{r}} \sim 600$
\cite{Gorshunov1999prb} and 1500 \cite{Travaglini1984prb}. Our
measurements then imply a reduced mass of the order of the bare
electron mass $m_\mathrm{e}$, suggesting that either electrons, holes,
or both are relatively light compared to reported values for the
effective mass in \smb{} of $m^*\sim
100\,m_\mathrm{e}$~\cite{Gorshunov1999prb}. Note that within this
model, the observed decrease in $\lambda$ near the surface is
primarily due to the absence of exitonic states and associated
magnetic fields in this region.

\mysection{Conclusions}
In conclusion, we observe fluctuating magnetic fields appearing only
below $\usim\SI{15}{K}$ in the bulk of \smb{}. Using \lem\
measurements we find that these fields are rapidly suppressed with
decreasing depth and probably disappear completely at the surface. We
attribute these fluctuating fields to excitonic states, whose extent
is limited to the bulk of \smb{} and disappears within
$\usim\SI{60}{nm}$ of its surface. An estimate of $\usim
0.01\mu_\mathrm{B}$ for the average magnitude of magnetic moments is
obtained from the distribution of fluctuating magnetic fields. We also
observe a slight increase in the distribution width of static magnetic
fields near the surface of \smb{}, hinting to the appearance of
additional magnetic moments in this region. Our results reveal a
complex magnetic behavior near the surface of the 3D topological Kondo
insulator \smb{}. We expect that the magnetic nature of the near
surface region of \smb{} may have significant implications on the
topological surface states at very low temperatures.

This work was performed at the Swiss Muon Source (S$\mu$S), Paul
Scherrer Institute (PSI, Switzerland). The work was supported, in
part, by the EPSRC, United Kingdom grant no.\ EP/I007210/1.


\begin{thebibliography}{49}%
\makeatletter
\providecommand \@ifxundefined [1]{%
 \@ifx{#1\undefined}
}%
\providecommand \@ifnum [1]{%
 \ifnum #1\expandafter \@firstoftwo
 \else \expandafter \@secondoftwo
 \fi
}%
\providecommand \@ifx [1]{%
 \ifx #1\expandafter \@firstoftwo
 \else \expandafter \@secondoftwo
 \fi
}%
\providecommand \natexlab [1]{#1}%
\providecommand \enquote  [1]{``#1''}%
\providecommand \bibnamefont  [1]{#1}%
\providecommand \bibfnamefont [1]{#1}%
\providecommand \citenamefont [1]{#1}%
\providecommand \href@noop [0]{\@secondoftwo}%
\providecommand \href [0]{\begingroup \@sanitize@url \@href}%
\providecommand \@href[1]{\@@startlink{#1}\@@href}%
\providecommand \@@href[1]{\endgroup#1\@@endlink}%
\providecommand \@sanitize@url [0]{\catcode `\\12\catcode `\$12\catcode
  `\&12\catcode `\#12\catcode `\^12\catcode `\_12\catcode `\%12\relax}%
\providecommand \@@startlink[1]{}%
\providecommand \@@endlink[0]{}%
\providecommand \url  [0]{\begingroup\@sanitize@url \@url }%
\providecommand \@url [1]{\endgroup\@href {#1}{\urlprefix }}%
\providecommand \urlprefix  [0]{URL }%
\providecommand \Eprint [0]{\href }%
\providecommand \doibase [0]{http://dx.doi.org/}%
\providecommand \selectlanguage [0]{\@gobble}%
\providecommand \bibinfo  [0]{\@secondoftwo}%
\providecommand \bibfield  [0]{\@secondoftwo}%
\providecommand \translation [1]{[#1]}%
\providecommand \BibitemOpen [0]{}%
\providecommand \bibitemStop [0]{}%
\providecommand \bibitemNoStop [0]{.\EOS\space}%
\providecommand \EOS [0]{\spacefactor3000\relax}%
\providecommand \BibitemShut  [1]{\csname bibitem#1\endcsname}%
\let\auto@bib@innerbib\@empty
\bibitem [{\citenamefont {Moore}(2010)}]{Moore2010n}%
  \BibitemOpen
  \bibfield  {author} {\bibinfo {author} {\bibfnamefont {J.~E.}\ \bibnamefont
  {Moore}},\ }\href {\doibase 10.1038/nature08916} {\bibfield  {journal}
  {\bibinfo  {journal} {Nature}\ }\textbf {\bibinfo {volume} {464}},\ \bibinfo
  {pages} {194} (\bibinfo {year} {2010})}\BibitemShut {NoStop}%
\bibitem [{\citenamefont {Dzero}\ \emph {et~al.}(2010)\citenamefont {Dzero},
  \citenamefont {Sun}, \citenamefont {Galitski},\ and\ \citenamefont
  {Coleman}}]{Dzero2010prl}%
  \BibitemOpen
  \bibfield  {author} {\bibinfo {author} {\bibfnamefont {M.}~\bibnamefont
  {Dzero}}, \bibinfo {author} {\bibfnamefont {K.}~\bibnamefont {Sun}}, \bibinfo
  {author} {\bibfnamefont {V.}~\bibnamefont {Galitski}}, \ and\ \bibinfo
  {author} {\bibfnamefont {P.}~\bibnamefont {Coleman}},\ }\href {\doibase
  10.1103/PhysRevLett.104.106408} {\bibfield  {journal} {\bibinfo  {journal}
  {Phys. Rev. Lett.}\ }\textbf {\bibinfo {volume} {104}},\ \bibinfo {pages}
  {106408} (\bibinfo {year} {2010})}\BibitemShut {NoStop}%
\bibitem [{\citenamefont {Alexandrov}\ \emph {et~al.}(2013)\citenamefont
  {Alexandrov}, \citenamefont {Dzero},\ and\ \citenamefont
  {Coleman}}]{Alexandrov2013prl}%
  \BibitemOpen
  \bibfield  {author} {\bibinfo {author} {\bibfnamefont {V.}~\bibnamefont
  {Alexandrov}}, \bibinfo {author} {\bibfnamefont {M.}~\bibnamefont {Dzero}}, \
  and\ \bibinfo {author} {\bibfnamefont {P.}~\bibnamefont {Coleman}},\ }\href
  {\doibase 10.1103/PhysRevLett.111.226403} {\bibfield  {journal} {\bibinfo
  {journal} {Phys. Rev. Lett.}\ }\textbf {\bibinfo {volume} {111}},\ \bibinfo
  {pages} {226403} (\bibinfo {year} {2013})}\BibitemShut {NoStop}%
\bibitem [{\citenamefont {Lu}\ \emph {et~al.}(2013)\citenamefont {Lu},
  \citenamefont {Zhao}, \citenamefont {Weng}, \citenamefont {Fang},\ and\
  \citenamefont {Dai}}]{Lu2013prl}%
  \BibitemOpen
  \bibfield  {author} {\bibinfo {author} {\bibfnamefont {F.}~\bibnamefont
  {Lu}}, \bibinfo {author} {\bibfnamefont {J.~Z.}~\bibnamefont {Zhao}}, \bibinfo
  {author} {\bibfnamefont {H.}~\bibnamefont {Weng}}, \bibinfo {author}
  {\bibfnamefont {Z.}~\bibnamefont {Fang}}, \ and\ \bibinfo {author}
  {\bibfnamefont {X.}~\bibnamefont {Dai}},\ }\href {\doibase
  10.1103/PhysRevLett.110.096401} {\bibfield  {journal} {\bibinfo  {journal}
  {Phys. Rev. Lett.}\ }\textbf {\bibinfo {volume} {110}},\ \bibinfo {pages}
  {096401} (\bibinfo {year} {2013})}\BibitemShut {NoStop}%
\bibitem [{\citenamefont {Dzero}\ \emph {et~al.}(2012)\citenamefont {Dzero},
  \citenamefont {Sun}, \citenamefont {Coleman},\ and\ \citenamefont
  {Galitski}}]{Dzero2012prb}%
  \BibitemOpen
  \bibfield  {author} {\bibinfo {author} {\bibfnamefont {M.}~\bibnamefont
  {Dzero}}, \bibinfo {author} {\bibfnamefont {K.}~\bibnamefont {Sun}}, \bibinfo
  {author} {\bibfnamefont {P.}~\bibnamefont {Coleman}}, \ and\ \bibinfo
  {author} {\bibfnamefont {V.}~\bibnamefont {Galitski}},\ }\href {\doibase
  10.1103/PhysRevB.85.045130} {\bibfield  {journal} {\bibinfo  {journal} {Phys.
  Rev. B}\ }\textbf {\bibinfo {volume} {85}},\ \bibinfo {pages} {045130}
  (\bibinfo {year} {2012})}\BibitemShut {NoStop}%
\bibitem [{\citenamefont {Wolgast}\ \emph {et~al.}(2013)\citenamefont
  {Wolgast}, \citenamefont {Kurdak}, \citenamefont {Sun}, \citenamefont
  {Allen}, \citenamefont {Kim},\ and\ \citenamefont {Fisk}}]{Wolgast2013prb}%
  \BibitemOpen
  \bibfield  {author} {\bibinfo {author} {\bibfnamefont {S.}~\bibnamefont
  {Wolgast}}, \bibinfo {author} {\bibfnamefont {{\c{C}}.}~\bibnamefont
  {Kurdak}}, \bibinfo {author} {\bibfnamefont {K.}~\bibnamefont {Sun}},
  \bibinfo {author} {\bibfnamefont {J.~W.}\ \bibnamefont {Allen}}, \bibinfo
  {author} {\bibfnamefont {D.-J.}\ \bibnamefont {Kim}}, \ and\ \bibinfo
  {author} {\bibfnamefont {Z.}~\bibnamefont {Fisk}},\ }\href {\doibase
  10.1103/PhysRevB.88.180405} {\bibfield  {journal} {\bibinfo  {journal} {Phys.
  Rev. B}\ }\textbf {\bibinfo {volume} {88}},\ \bibinfo {pages} {180405}
  (\bibinfo {year} {2013})}\BibitemShut {NoStop}%
\bibitem [{\citenamefont {Kim}\ \emph {et~al.}(2013)\citenamefont {Kim},
  \citenamefont {Thomas}, \citenamefont {Grant}, \citenamefont {Botimer},
  \citenamefont {Fisk},\ and\ \citenamefont {Xia}}]{Kim2013sr}%
  \BibitemOpen
  \bibfield  {author} {\bibinfo {author} {\bibfnamefont {D.~J.}\ \bibnamefont
  {Kim}}, \bibinfo {author} {\bibfnamefont {S.}~\bibnamefont {Thomas}},
  \bibinfo {author} {\bibfnamefont {T.}~\bibnamefont {Grant}}, \bibinfo
  {author} {\bibfnamefont {J.}~\bibnamefont {Botimer}}, \bibinfo {author}
  {\bibfnamefont {Z.}~\bibnamefont {Fisk}}, \ and\ \bibinfo {author}
  {\bibfnamefont {J.}~\bibnamefont {Xia}},\ }\href {\doibase 10.1038/srep03150}
  {\bibfield  {journal} {\bibinfo  {journal} {Scientific Reports}\ }\textbf
  {\bibinfo {volume} {3}},\ \bibinfo {pages} {3150} (\bibinfo {year}
  {2013})}\BibitemShut {NoStop}%
\bibitem [{\citenamefont {Zhang}\ \emph {et~al.}(2013)\citenamefont {Zhang},
  \citenamefont {Butch}, \citenamefont {Syers}, \citenamefont {Ziemak},
  \citenamefont {Greene},\ and\ \citenamefont {Paglione}}]{Zhang2013prx}%
  \BibitemOpen
  \bibfield  {author} {\bibinfo {author} {\bibfnamefont {X.}~\bibnamefont
  {Zhang}}, \bibinfo {author} {\bibfnamefont {N.~P.}\ \bibnamefont {Butch}},
  \bibinfo {author} {\bibfnamefont {P.}~\bibnamefont {Syers}}, \bibinfo
  {author} {\bibfnamefont {S.}~\bibnamefont {Ziemak}}, \bibinfo {author}
  {\bibfnamefont {R.~L.}\ \bibnamefont {Greene}}, \ and\ \bibinfo {author}
  {\bibfnamefont {J.}~\bibnamefont {Paglione}},\ }\href {\doibase
  10.1103/PhysRevX.3.011011} {\bibfield  {journal} {\bibinfo  {journal} {Phys.
  Rev. X}\ }\textbf {\bibinfo {volume} {3}},\ \bibinfo {pages} {011011}
  (\bibinfo {year} {2013})}\BibitemShut {NoStop}%
\bibitem [{\citenamefont {Kim}\ \emph {et~al.}(2014)\citenamefont {Kim},
  \citenamefont {Xia},\ and\ \citenamefont {Fisk}}]{Kim2014nm}%
  \BibitemOpen
  \bibfield  {author} {\bibinfo {author} {\bibfnamefont {D.~J.}\ \bibnamefont
  {Kim}}, \bibinfo {author} {\bibfnamefont {J.}~\bibnamefont {Xia}}, \ and\
  \bibinfo {author} {\bibfnamefont {Z.}~\bibnamefont {Fisk}},\ }\href {\doibase
  10.1038/nmat3913} {\bibfield  {journal} {\bibinfo  {journal} {Nat Mater}\
  }\textbf {\bibinfo {volume} {13}},\ \bibinfo {pages} {466} (\bibinfo {year}
  {2014})}\BibitemShut {NoStop}%
\bibitem [{\citenamefont {Xu}\ \emph {et~al.}(2013)\citenamefont {Xu},
  \citenamefont {Shi}, \citenamefont {Biswas}, \citenamefont {Matt},
  \citenamefont {Dhaka}, \citenamefont {Huang}, \citenamefont {Plumb},
  \citenamefont {Radovi\'c}, \citenamefont {Dil}, \citenamefont {Pomjakushina},
  \citenamefont {Conder}, \citenamefont {Amato}, \citenamefont {Salman},
  \citenamefont {Paul}, \citenamefont {Mesot}, \citenamefont {Ding},\ and\
  \citenamefont {Shi}}]{Xu2013prb}%
  \BibitemOpen
  \bibfield  {author} {\bibinfo {author} {\bibfnamefont {N.}~\bibnamefont
  {Xu}}, \bibinfo {author} {\bibfnamefont {X.}~\bibnamefont {Shi}}, \bibinfo
  {author} {\bibfnamefont {P.~K.}\ \bibnamefont {Biswas}}, \bibinfo {author}
  {\bibfnamefont {C.~E.}\ \bibnamefont {Matt}}, \bibinfo {author}
  {\bibfnamefont {R.~S.}\ \bibnamefont {Dhaka}}, \bibinfo {author}
  {\bibfnamefont {Y.}~\bibnamefont {Huang}}, \bibinfo {author} {\bibfnamefont
  {N.~C.}\ \bibnamefont {Plumb}}, \bibinfo {author} {\bibfnamefont
  {M.}~\bibnamefont {Radovi\'c}}, \bibinfo {author} {\bibfnamefont {J.~H.}\
  \bibnamefont {Dil}}, \bibinfo {author} {\bibfnamefont {E.}~\bibnamefont
  {Pomjakushina}}, \bibinfo {author} {\bibfnamefont {K.}~\bibnamefont
  {Conder}}, \bibinfo {author} {\bibfnamefont {A.}~\bibnamefont {Amato}},
  \bibinfo {author} {\bibfnamefont {Z.}~\bibnamefont {Salman}}, \bibinfo
  {author} {\bibfnamefont {D.~M.}\ \bibnamefont {Paul}}, \bibinfo {author}
  {\bibfnamefont {J.}~\bibnamefont {Mesot}}, \bibinfo {author} {\bibfnamefont
  {H.}~\bibnamefont {Ding}}, \ and\ \bibinfo {author} {\bibfnamefont
  {M.}~\bibnamefont {Shi}},\ }\href {\doibase 10.1103/PhysRevB.88.121102}
  {\bibfield  {journal} {\bibinfo  {journal} {Phys. Rev. B}\ }\textbf {\bibinfo
  {volume} {88}},\ \bibinfo {pages} {121102} (\bibinfo {year}
  {2013})}\BibitemShut {NoStop}%
\bibitem [{\citenamefont {Neupane}\ \emph {et~al.}(2013)\citenamefont
  {Neupane}, \citenamefont {Alidoust}, \citenamefont {Xu}, \citenamefont
  {Kondo}, \citenamefont {Ishida}, \citenamefont {Kim}, \citenamefont {Liu},
  \citenamefont {Belopolski}, \citenamefont {Jo}, \citenamefont {Chang},
  \citenamefont {Jeng}, \citenamefont {Durakiewicz}, \citenamefont {Balicas},
  \citenamefont {Lin}, \citenamefont {Bansil}, \citenamefont {Shin},
  \citenamefont {Fisk},\ and\ \citenamefont {Hasan}}]{Neupane2013nc}%
  \BibitemOpen
  \bibfield  {author} {\bibinfo {author} {\bibfnamefont {M.}~\bibnamefont
  {Neupane}}, \bibinfo {author} {\bibfnamefont {N.}~\bibnamefont {Alidoust}},
  \bibinfo {author} {\bibfnamefont {S.-Y.}\ \bibnamefont {Xu}}, \bibinfo
  {author} {\bibfnamefont {T.}~\bibnamefont {Kondo}}, \bibinfo {author}
  {\bibfnamefont {Y.}~\bibnamefont {Ishida}}, \bibinfo {author} {\bibfnamefont
  {D.~J.}\ \bibnamefont {Kim}}, \bibinfo {author} {\bibfnamefont
  {C.}~\bibnamefont {Liu}}, \bibinfo {author} {\bibfnamefont {I.}~\bibnamefont
  {Belopolski}}, \bibinfo {author} {\bibfnamefont {Y.~J.}\ \bibnamefont {Jo}},
  \bibinfo {author} {\bibfnamefont {T.-R.}\ \bibnamefont {Chang}}, \bibinfo
  {author} {\bibfnamefont {H.-T.}\ \bibnamefont {Jeng}}, \bibinfo {author}
  {\bibfnamefont {T.}~\bibnamefont {Durakiewicz}}, \bibinfo {author}
  {\bibfnamefont {L.}~\bibnamefont {Balicas}}, \bibinfo {author} {\bibfnamefont
  {H.}~\bibnamefont {Lin}}, \bibinfo {author} {\bibfnamefont {A.}~\bibnamefont
  {Bansil}}, \bibinfo {author} {\bibfnamefont {S.}~\bibnamefont {Shin}},
  \bibinfo {author} {\bibfnamefont {Z.}~\bibnamefont {Fisk}}, \ and\ \bibinfo
  {author} {\bibfnamefont {M.~Z.}\ \bibnamefont {Hasan}},\ }\href {\doibase
  10.1038/ncomms3991} {\bibfield  {journal} {\bibinfo  {journal} {Nat.
  Commun.}\ }\textbf {\bibinfo {volume} {4}},\ \bibinfo {pages} {2991}
  (\bibinfo {year} {2013})}\BibitemShut {NoStop}%
\bibitem [{\citenamefont {Jiang}\ \emph {et~al.}(2013)\citenamefont {Jiang},
  \citenamefont {Li}, \citenamefont {Zhang}, \citenamefont {Sun}, \citenamefont
  {Chen}, \citenamefont {Ye}, \citenamefont {Xu}, \citenamefont {Ge},
  \citenamefont {Tan}, \citenamefont {Niu}, \citenamefont {Xia}, \citenamefont
  {Xie}, \citenamefont {Li}, \citenamefont {Chen}, \citenamefont {Wen},\ and\
  \citenamefont {Feng}}]{Jiang2013nc}%
  \BibitemOpen
  \bibfield  {author} {\bibinfo {author} {\bibfnamefont {J.}~\bibnamefont
  {Jiang}}, \bibinfo {author} {\bibfnamefont {S.}~\bibnamefont {Li}}, \bibinfo
  {author} {\bibfnamefont {T.}~\bibnamefont {Zhang}}, \bibinfo {author}
  {\bibfnamefont {Z.}~\bibnamefont {Sun}}, \bibinfo {author} {\bibfnamefont
  {F.}~\bibnamefont {Chen}}, \bibinfo {author} {\bibfnamefont {Z.~R.}\
  \bibnamefont {Ye}}, \bibinfo {author} {\bibfnamefont {M.}~\bibnamefont {Xu}},
  \bibinfo {author} {\bibfnamefont {Q.~Q.}\ \bibnamefont {Ge}}, \bibinfo
  {author} {\bibfnamefont {S.~Y.}\ \bibnamefont {Tan}}, \bibinfo {author}
  {\bibfnamefont {X.~H.}\ \bibnamefont {Niu}}, \bibinfo {author} {\bibfnamefont
  {M.}~\bibnamefont {Xia}}, \bibinfo {author} {\bibfnamefont {B.~P.}\
  \bibnamefont {Xie}}, \bibinfo {author} {\bibfnamefont {Y.~F.}\ \bibnamefont
  {Li}}, \bibinfo {author} {\bibfnamefont {X.~H.}\ \bibnamefont {Chen}},
  \bibinfo {author} {\bibfnamefont {H.~H.}\ \bibnamefont {Wen}}, \ and\
  \bibinfo {author} {\bibfnamefont {D.~L.}\ \bibnamefont {Feng}},\ }\href
  {\doibase 10.1038/ncomms4010} {\bibfield  {journal} {\bibinfo  {journal}
  {Nat. Commun.}\ }\textbf {\bibinfo {volume} {4}},\ \bibinfo {pages} {3010}
  (\bibinfo {year} {2013})}\BibitemShut {NoStop}%
\bibitem [{\citenamefont {Xu}\ \emph {et~al.}(2014)\citenamefont {Xu},
  \citenamefont {Biswas}, \citenamefont {Dil}, \citenamefont {Dhaka},
  \citenamefont {Landolt}, \citenamefont {Muff}, \citenamefont {Matt},
  \citenamefont {Shi}, \citenamefont {Plumb}, \citenamefont {Radovi\'c},
  \citenamefont {Pomjakushina}, \citenamefont {Conder}, \citenamefont {Amato},
  \citenamefont {Borisenko}, \citenamefont {Yu}, \citenamefont {Weng},
  \citenamefont {Fang}, \citenamefont {Dai}, \citenamefont {Mesot},
  \citenamefont {Ding},\ and\ \citenamefont {Shi}}]{Xu2014nc}%
  \BibitemOpen
  \bibfield  {author} {\bibinfo {author} {\bibfnamefont {N.}~\bibnamefont
  {Xu}}, \bibinfo {author} {\bibfnamefont {P.~K.}\ \bibnamefont {Biswas}},
  \bibinfo {author} {\bibfnamefont {J.~H.}\ \bibnamefont {Dil}}, \bibinfo
  {author} {\bibfnamefont {R.~S.}\ \bibnamefont {Dhaka}}, \bibinfo {author}
  {\bibfnamefont {G.}~\bibnamefont {Landolt}}, \bibinfo {author} {\bibfnamefont
  {S.}~\bibnamefont {Muff}}, \bibinfo {author} {\bibfnamefont {C.~E.}\
  \bibnamefont {Matt}}, \bibinfo {author} {\bibfnamefont {X.}~\bibnamefont
  {Shi}}, \bibinfo {author} {\bibfnamefont {N.~C.}\ \bibnamefont {Plumb}},
  \bibinfo {author} {\bibfnamefont {M.}~\bibnamefont {Radovi\'c}}, \bibinfo
  {author} {\bibfnamefont {E.}~\bibnamefont {Pomjakushina}}, \bibinfo {author}
  {\bibfnamefont {K.}~\bibnamefont {Conder}}, \bibinfo {author} {\bibfnamefont
  {A.}~\bibnamefont {Amato}}, \bibinfo {author} {\bibfnamefont {S.~V.}\
  \bibnamefont {Borisenko}}, \bibinfo {author} {\bibfnamefont {R.}~\bibnamefont
  {Yu}}, \bibinfo {author} {\bibfnamefont {H.-M.}\ \bibnamefont {Weng}},
  \bibinfo {author} {\bibfnamefont {Z.}~\bibnamefont {Fang}}, \bibinfo {author}
  {\bibfnamefont {X.}~\bibnamefont {Dai}}, \bibinfo {author} {\bibfnamefont
  {J.}~\bibnamefont {Mesot}}, \bibinfo {author} {\bibfnamefont
  {H.}~\bibnamefont {Ding}}, \ and\ \bibinfo {author} {\bibfnamefont
  {M.}~\bibnamefont {Shi}},\ }\href {\doibase 10.1038/ncomms5566} {\bibfield
  {journal} {\bibinfo  {journal} {Nat. Commun.}\ }\textbf {\bibinfo {volume}
  {5}},\ \bibinfo {pages} {4566} (\bibinfo {year} {2014})}\BibitemShut
  {NoStop}%
\bibitem [{\citenamefont {Aeppli}\ and\ \citenamefont
  {Fisk}(1992)}]{Aeppli1992ccmp}%
  \BibitemOpen
  \bibfield  {author} {\bibinfo {author} {\bibfnamefont {G.}~\bibnamefont
  {Aeppli}}\ and\ \bibinfo {author} {\bibfnamefont {Z.}~\bibnamefont {Fisk}},\
  }\href@noop {} {\bibfield  {journal} {\bibinfo  {journal} {Comments Condens.
  Matter Phys.}\ }\textbf {\bibinfo {volume} {16}},\ \bibinfo {pages} {155}
  (\bibinfo {year} {1992})}\BibitemShut {NoStop}%
\bibitem [{\citenamefont {Riseborough}(2000)}]{Riseborough2000adp}%
  \BibitemOpen
  \bibfield  {author} {\bibinfo {author} {\bibfnamefont {P.~S.}\ \bibnamefont
  {Riseborough}},\ }\href {\doibase
  10.1002/1521-3889(200010)9:9/10<813::AID-ANDP813>3.0.CO;2-E} {\bibfield
  {journal} {\bibinfo  {journal} {Annalen der Physik}\ }\textbf {\bibinfo
  {volume} {9}},\ \bibinfo {pages} {813} (\bibinfo {year} {2000})}\BibitemShut
  {NoStop}%
\bibitem [{\citenamefont {Coleman}(2007)}]{Coleman2007}%
  \BibitemOpen
  \bibfield  {author} {\bibinfo {author} {\bibfnamefont {P.}~\bibnamefont
  {Coleman}},\ }\href
  {http://onlinelibrary.wiley.com/doi/10.1002/9780470022184.hmm105/full} {\emph
  {\bibinfo {title} {Heavy Fermions: electrons at the edge of magnetism}}}\
  (\bibinfo  {publisher} {Wiley Online Library},\ \bibinfo {year}
  {2007})\BibitemShut {NoStop}%
\bibitem [{\citenamefont {Menth}\ \emph {et~al.}(1969)\citenamefont {Menth},
  \citenamefont {Buehler},\ and\ \citenamefont {Geballe}}]{Menth1969prl}%
  \BibitemOpen
  \bibfield  {author} {\bibinfo {author} {\bibfnamefont {A.}~\bibnamefont
  {Menth}}, \bibinfo {author} {\bibfnamefont {E.}~\bibnamefont {Buehler}}, \
  and\ \bibinfo {author} {\bibfnamefont {T.~H.}\ \bibnamefont {Geballe}},\
  }\href {\doibase 10.1103/PhysRevLett.22.295} {\bibfield  {journal} {\bibinfo
  {journal} {Phys. Rev. Lett.}\ }\textbf {\bibinfo {volume} {22}},\ \bibinfo
  {pages} {295} (\bibinfo {year} {1969})}\BibitemShut {NoStop}%
\bibitem [{\citenamefont {Allen}\ \emph {et~al.}(1979)\citenamefont {Allen},
  \citenamefont {Batlogg},\ and\ \citenamefont {Wachter}}]{Allen1979PRB}%
  \BibitemOpen
  \bibfield  {author} {\bibinfo {author} {\bibfnamefont {J.~W.}\ \bibnamefont
  {Allen}}, \bibinfo {author} {\bibfnamefont {B.}~\bibnamefont {Batlogg}}, \
  and\ \bibinfo {author} {\bibfnamefont {P.}~\bibnamefont {Wachter}},\ }\href
  {\doibase 10.1103/PhysRevB.20.4807} {\bibfield  {journal} {\bibinfo
  {journal} {Phys. Rev. B}\ }\textbf {\bibinfo {volume} {20}},\ \bibinfo
  {pages} {4807} (\bibinfo {year} {1979})}\BibitemShut {NoStop}%
\bibitem [{\citenamefont {Cooley}\ \emph {et~al.}(1995)\citenamefont {Cooley},
  \citenamefont {Aronson}, \citenamefont {Fisk},\ and\ \citenamefont
  {Canfield}}]{Cooley1995prl}%
  \BibitemOpen
  \bibfield  {author} {\bibinfo {author} {\bibfnamefont {J.~C.}\ \bibnamefont
  {Cooley}}, \bibinfo {author} {\bibfnamefont {M.~C.}\ \bibnamefont {Aronson}},
  \bibinfo {author} {\bibfnamefont {Z.}~\bibnamefont {Fisk}}, \ and\ \bibinfo
  {author} {\bibfnamefont {P.~C.}\ \bibnamefont {Canfield}},\ }\href {\doibase
  10.1103/PhysRevLett.74.1629} {\bibfield  {journal} {\bibinfo  {journal}
  {Phys. Rev. Lett.}\ }\textbf {\bibinfo {volume} {74}},\ \bibinfo {pages}
  {1629} (\bibinfo {year} {1995})}\BibitemShut {NoStop}%
\bibitem [{\citenamefont {Miyazaki}\ \emph {et~al.}(2012)\citenamefont
  {Miyazaki}, \citenamefont {Hajiri}, \citenamefont {Ito}, \citenamefont
  {Kunii},\ and\ \citenamefont {Kimura}}]{Miyazaki2012prb}%
  \BibitemOpen
  \bibfield  {author} {\bibinfo {author} {\bibfnamefont {H.}~\bibnamefont
  {Miyazaki}}, \bibinfo {author} {\bibfnamefont {T.}~\bibnamefont {Hajiri}},
  \bibinfo {author} {\bibfnamefont {T.}~\bibnamefont {Ito}}, \bibinfo {author}
  {\bibfnamefont {S.}~\bibnamefont {Kunii}}, \ and\ \bibinfo {author}
  {\bibfnamefont {S.-I.}~\bibnamefont {Kimura}},\ }\href {\doibase
  10.1103/PhysRevB.86.075105} {\bibfield  {journal} {\bibinfo  {journal} {Phys.
  Rev. B}\ }\textbf {\bibinfo {volume} {86}},\ \bibinfo {pages} {075105}
  (\bibinfo {year} {2012})}\BibitemShut {NoStop}%
\bibitem [{\citenamefont {Denlinger}\ \emph {et~al.}(2000)\citenamefont
  {Denlinger}, \citenamefont {Gweon}, \citenamefont {Allen}, \citenamefont
  {Olson}, \citenamefont {Dalichaouch}, \citenamefont {Lee}, \citenamefont
  {Maple}, \citenamefont {Fisk}, \citenamefont {Canfield},\ and\ \citenamefont
  {Armstrong}}]{Denlinger2000pb}%
  \BibitemOpen
  \bibfield  {author} {\bibinfo {author} {\bibfnamefont {J.~D.}\ \bibnamefont
  {Denlinger}}, \bibinfo {author} {\bibfnamefont {G.~H.}\ \bibnamefont
  {Gweon}}, \bibinfo {author} {\bibfnamefont {J.~W.}\ \bibnamefont {Allen}},
  \bibinfo {author} {\bibfnamefont {C.~G.}\ \bibnamefont {Olson}}, \bibinfo
  {author} {\bibfnamefont {Y.}~\bibnamefont {Dalichaouch}}, \bibinfo {author}
  {\bibfnamefont {B.~W.}\ \bibnamefont {Lee}}, \bibinfo {author} {\bibfnamefont
  {M.~B.}\ \bibnamefont {Maple}}, \bibinfo {author} {\bibfnamefont
  {Z.}~\bibnamefont {Fisk}}, \bibinfo {author} {\bibfnamefont {P.~C.}\
  \bibnamefont {Canfield}}, \ and\ \bibinfo {author} {\bibfnamefont {P.~E.}\
  \bibnamefont {Armstrong}},\ }\href {\doibase 10.1016/S0921-4526(99)00915-1}
  {\bibfield  {journal} {\bibinfo  {journal} {Physica B: Condensed Matter}\
  }\textbf {\bibinfo {volume} {281-282}},\ \bibinfo {pages} {716} (\bibinfo
  {year} {2000})}\BibitemShut {NoStop}%
\bibitem [{\citenamefont {Alekseev}\ \emph {et~al.}(1993)\citenamefont
  {Alekseev}, \citenamefont {Mignot}, \citenamefont {Rossat-Mignod},
  \citenamefont {Lazukov},\ and\ \citenamefont {Sadikov}}]{Alekseev1993pb}%
  \BibitemOpen
  \bibfield  {author} {\bibinfo {author} {\bibfnamefont {P.}~\bibnamefont
  {Alekseev}}, \bibinfo {author} {\bibfnamefont {J.-M.}\ \bibnamefont
  {Mignot}}, \bibinfo {author} {\bibfnamefont {J.}~\bibnamefont
  {Rossat-Mignod}}, \bibinfo {author} {\bibfnamefont {V.}~\bibnamefont
  {Lazukov}}, \ and\ \bibinfo {author} {\bibfnamefont {I.}~\bibnamefont
  {Sadikov}},\ }\href {\doibase 10.1016/0921-4526(93)90580-Y} {\bibfield
  {journal} {\bibinfo  {journal} {Physica B: Condensed Matter}\ }\textbf
  {\bibinfo {volume} {186-188}},\ \bibinfo {pages} {384} (\bibinfo {year}
  {1993})}\BibitemShut {NoStop}%
\bibitem [{\citenamefont {Denlinger}\ \emph {et~al.}(2014)\citenamefont
  {Denlinger}, \citenamefont {Allen}, \citenamefont {Kang}, \citenamefont
  {Sun}, \citenamefont {Min}, \citenamefont {Kim},\ and\ \citenamefont
  {Fisk}}]{Denlinger2014jpscs}%
  \BibitemOpen
  \bibfield  {author} {\bibinfo {author} {\bibfnamefont {J.~D.}\ \bibnamefont
  {Denlinger}}, \bibinfo {author} {\bibfnamefont {J.~W.}\ \bibnamefont
  {Allen}}, \bibinfo {author} {\bibfnamefont {J.-S.}\ \bibnamefont {Kang}},
  \bibinfo {author} {\bibfnamefont {K.}~\bibnamefont {Sun}}, \bibinfo {author}
  {\bibfnamefont {B.-I.}\ \bibnamefont {Min}}, \bibinfo {author} {\bibfnamefont
  {D.-J.}\ \bibnamefont {Kim}}, \ and\ \bibinfo {author} {\bibfnamefont
  {Z.}~\bibnamefont {Fisk}},\ }\href {\doibase 10.7566/JPSCP.3.017038}
  {\bibfield  {journal} {\bibinfo  {journal} {JPS Conf. Proc.}\ }\textbf
  {\bibinfo {volume} {3}},\ \bibinfo {pages} {017038} (\bibinfo {year}
  {2014})}\BibitemShut {NoStop}%
\bibitem [{\citenamefont {Denlinger}\ \emph {et~al.}()\citenamefont
  {Denlinger}, \citenamefont {Allen}, \citenamefont {Kang}, \citenamefont
  {Sun}, \citenamefont {Kim}, \citenamefont {Shim}, \citenamefont {Min},
  \citenamefont {Kim},\ and\ \citenamefont {Fisk}}]{Denlinger2013arxiv}%
  \BibitemOpen
  \bibfield  {author} {\bibinfo {author} {\bibfnamefont {J.~D.}\ \bibnamefont
  {Denlinger}}, \bibinfo {author} {\bibfnamefont {J.~W.}\ \bibnamefont
  {Allen}}, \bibinfo {author} {\bibfnamefont {J.-S.}\ \bibnamefont {Kang}},
  \bibinfo {author} {\bibfnamefont {K.}~\bibnamefont {Sun}}, \bibinfo {author}
  {\bibfnamefont {J.-W.}\ \bibnamefont {Kim}}, \bibinfo {author} {\bibfnamefont
  {J.~H.}\ \bibnamefont {Shim}}, \bibinfo {author} {\bibfnamefont {B.~I.}\
  \bibnamefont {Min}}, \bibinfo {author} {\bibfnamefont {D.-J.}\ \bibnamefont
  {Kim}}, \ and\ \bibinfo {author} {\bibfnamefont {Z.}~\bibnamefont {Fisk}},\
  }\href {http://arxiv.org/abs/1312.6637} {\bibinfo  {journal}
  {{arXiv}:1312.6637}\ }\BibitemShut {NoStop}%
\bibitem [{\citenamefont {Alekseev}\ \emph {et~al.}(1995)\citenamefont
  {Alekseev}, \citenamefont {Mignot}, \citenamefont {Rossat-Mignod},
  \citenamefont {Lazukov}, \citenamefont {Sadikov}, \citenamefont
  {Konovalova},\ and\ \citenamefont {Paderno}}]{Alekseev1995jpcm}%
  \BibitemOpen
\bibfield  {journal} {  }\bibfield  {author} {\bibinfo {author} {\bibfnamefont
  {P.~A.}\ \bibnamefont {Alekseev}}, \bibinfo {author} {\bibfnamefont {J.~M.}\
  \bibnamefont {Mignot}}, \bibinfo {author} {\bibfnamefont {J.}~\bibnamefont
  {Rossat-Mignod}}, \bibinfo {author} {\bibfnamefont {V.~N.}\ \bibnamefont
  {Lazukov}}, \bibinfo {author} {\bibfnamefont {I.~P.}\ \bibnamefont
  {Sadikov}}, \bibinfo {author} {\bibfnamefont {E.~S.}\ \bibnamefont
  {Konovalova}}, \ and\ \bibinfo {author} {\bibfnamefont {Y.~B.}\ \bibnamefont
  {Paderno}},\ }\href {\doibase 10.1088/0953-8984/7/2/007} {\bibfield
  {journal} {\bibinfo  {journal} {J. Phys.: Condens. Matter}\ }\textbf
  {\bibinfo {volume} {7}},\ \bibinfo {pages} {289} (\bibinfo {year}
  {1995})}\BibitemShut {NoStop}%
\bibitem [{\citenamefont {Alekseev}\ \emph {et~al.}(2010)\citenamefont
  {Alekseev}, \citenamefont {Lazukov}, \citenamefont {Nemkovskii},\ and\
  \citenamefont {Sadikov}}]{Alekseev2010jetp}%
  \BibitemOpen
  \bibfield  {author} {\bibinfo {author} {\bibfnamefont {P.~A.}\ \bibnamefont
  {Alekseev}}, \bibinfo {author} {\bibfnamefont {V.~N.}\ \bibnamefont
  {Lazukov}}, \bibinfo {author} {\bibfnamefont {K.~S.}\ \bibnamefont
  {Nemkovskii}}, \ and\ \bibinfo {author} {\bibfnamefont {I.~P.}\ \bibnamefont
  {Sadikov}},\ }\href {\doibase 10.1134/S1063776110080224} {\bibfield
  {journal} {\bibinfo  {journal} {J. Exp. Theor. Phys.}\ }\textbf {\bibinfo
  {volume} {111}},\ \bibinfo {pages} {285} (\bibinfo {year}
  {2010})}\BibitemShut {NoStop}%
\bibitem [{\citenamefont {Fuhrman}\ \emph {et~al.}(2015)\citenamefont
  {Fuhrman}, \citenamefont {Leiner}, \citenamefont {Nikoli\'c}, \citenamefont
  {Granroth}, \citenamefont {Stone}, \citenamefont {Lumsden}, \citenamefont
  {{DeBeer}-Schmitt}, \citenamefont {Alekseev}, \citenamefont {Mignot},
  \citenamefont {Koohpayeh}, \citenamefont {Cottingham}, \citenamefont
  {Phelan}, \citenamefont {Schoop}, \citenamefont {{McQueen}},\ and\
  \citenamefont {Broholm}}]{Fuhrman2015prl}%
  \BibitemOpen
  \bibfield  {author} {\bibinfo {author} {\bibfnamefont {W.~T.}~\bibnamefont
  {Fuhrman}}, \bibinfo {author} {\bibfnamefont {J.}~\bibnamefont {Leiner}},
  \bibinfo {author} {\bibfnamefont {P.}~\bibnamefont {Nikoli\'c}}, \bibinfo
  {author} {\bibfnamefont {G.~E.}~\bibnamefont {Granroth}}, \bibinfo {author}
  {\bibfnamefont {M.~B.}~\bibnamefont {Stone}}, \bibinfo {author} {\bibfnamefont
  {M.~D.}~\bibnamefont {Lumsden}}, \bibinfo {author} {\bibfnamefont
  {L.}~\bibnamefont {{DeBeer}-Schmitt}}, \bibinfo {author} {\bibfnamefont
  {P.~A.}~\bibnamefont {Alekseev}}, \bibinfo {author} {\bibfnamefont {J.-M.}\
  \bibnamefont {Mignot}}, \bibinfo {author} {\bibfnamefont {S.~M.}~\bibnamefont
  {Koohpayeh}}, \bibinfo {author} {\bibfnamefont {P.}~\bibnamefont
  {Cottingham}}, \bibinfo {author} {\bibfnamefont {W.~A.}\ \bibnamefont
  {Phelan}}, \bibinfo {author} {\bibfnamefont {L.}~\bibnamefont {Schoop}},
  \bibinfo {author} {\bibfnamefont {T.~M.}~\bibnamefont {{McQueen}}}, \ and\
  \bibinfo {author} {\bibfnamefont {C.}~\bibnamefont {Broholm}},\ }\href
  {\doibase 10.1103/PhysRevLett.114.036401} {\bibfield  {journal} {\bibinfo
  {journal} {Phys. Rev. Lett.}\ }\textbf {\bibinfo {volume} {114}},\ \bibinfo
  {pages} {036401} (\bibinfo {year} {2015})}\BibitemShut {NoStop}%
\bibitem [{\citenamefont {Takigawa}\ \emph {et~al.}(1981)\citenamefont
  {Takigawa}, \citenamefont {Yasuoka}, \citenamefont {Kitaoka}, \citenamefont
  {Tanaka}, \citenamefont {Nozaki},\ and\ \citenamefont
  {Ishizawa}}]{Takigawa1981jpsj}%
  \BibitemOpen
  \bibfield  {author} {\bibinfo {author} {\bibfnamefont {M.}~\bibnamefont
  {Takigawa}}, \bibinfo {author} {\bibfnamefont {H.}~\bibnamefont {Yasuoka}},
  \bibinfo {author} {\bibfnamefont {Y.}~\bibnamefont {Kitaoka}}, \bibinfo
  {author} {\bibfnamefont {T.}~\bibnamefont {Tanaka}}, \bibinfo {author}
  {\bibfnamefont {H.}~\bibnamefont {Nozaki}}, \ and\ \bibinfo {author}
  {\bibfnamefont {Y.}~\bibnamefont {Ishizawa}},\ }\href {\doibase
  10.1143/JPSJ.50.2525} {\bibfield  {journal} {\bibinfo  {journal} {Journal of
  the Physical Society of Japan}\ }\textbf {\bibinfo {volume} {50}},\ \bibinfo
  {pages} {2525} (\bibinfo {year} {1981})}\BibitemShut {NoStop}%
\bibitem [{\citenamefont {Caldwell}(2004)}]{Caldwell2004}%
  \BibitemOpen
  \bibfield  {author} {\bibinfo {author} {\bibfnamefont {T.}~\bibnamefont
  {Caldwell}},\ }\emph {\bibinfo {title} {{Nuclear} Magnetic Resonance Studies
  of Field Effects on Single Crystal {SmB}$_6$}},\ \href
  {http://etd.lib.fsu.edu/theses/available/etd-03252004-143748/} {Ph.D.
  thesis},\ \bibinfo  {school} {The Florida State University} (\bibinfo {year}
  {2004})\BibitemShut {NoStop}%
\bibitem [{\citenamefont {Caldwell}\ \emph {et~al.}(2007)\citenamefont
  {Caldwell}, \citenamefont {Reyes}, \citenamefont {Moulton}, \citenamefont
  {Kuhns}, \citenamefont {Hoch}, \citenamefont {Schlottmann},\ and\
  \citenamefont {Fisk}}]{Caldwell2007prb}%
  \BibitemOpen
  \bibfield  {author} {\bibinfo {author} {\bibfnamefont {T.}~\bibnamefont
  {Caldwell}}, \bibinfo {author} {\bibfnamefont {A.~P.}\ \bibnamefont {Reyes}},
  \bibinfo {author} {\bibfnamefont {W.~G.}\ \bibnamefont {Moulton}}, \bibinfo
  {author} {\bibfnamefont {P.~L.}\ \bibnamefont {Kuhns}}, \bibinfo {author}
  {\bibfnamefont {M.~J.~R.}\ \bibnamefont {Hoch}}, \bibinfo {author}
  {\bibfnamefont {P.}~\bibnamefont {Schlottmann}}, \ and\ \bibinfo {author}
  {\bibfnamefont {Z.}~\bibnamefont {Fisk}},\ }\href {\doibase
  10.1103/PhysRevB.75.075106} {\bibfield  {journal} {\bibinfo  {journal} {Phys.
  Rev. B}\ }\textbf {\bibinfo {volume} {75}},\ \bibinfo {pages} {075106}
  (\bibinfo {year} {2007})}\BibitemShut {NoStop}%
\bibitem [{\citenamefont {Schlottmann}(2014)}]{Schlottmann2014prb}%
  \BibitemOpen
  \bibfield  {author} {\bibinfo {author} {\bibfnamefont {P.}~\bibnamefont
  {Schlottmann}},\ }\href {\doibase 10.1103/PhysRevB.90.165127} {\bibfield
  {journal} {\bibinfo  {journal} {Phys. Rev. B}\ }\textbf {\bibinfo {volume}
  {90}},\ \bibinfo {pages} {165127} (\bibinfo {year} {2014})}\BibitemShut
  {NoStop}%
\bibitem [{\citenamefont {Biswas}\ \emph {et~al.}(2014)\citenamefont {Biswas},
  \citenamefont {Salman}, \citenamefont {Neupert}, \citenamefont {Morenzoni},
  \citenamefont {Pomjakushina}, \citenamefont {von Rohr}, \citenamefont
  {Conder}, \citenamefont {Balakrishnan}, \citenamefont {Hatnean},
  \citenamefont {Lees}, \citenamefont {Paul}, \citenamefont {Schilling},
  \citenamefont {Baines}, \citenamefont {Luetkens}, \citenamefont {Khasanov},\
  and\ \citenamefont {Amato}}]{Biswas2014prb}%
  \BibitemOpen
  \bibfield  {author} {\bibinfo {author} {\bibfnamefont {P.~K.}\ \bibnamefont
  {Biswas}}, \bibinfo {author} {\bibfnamefont {Z.}~\bibnamefont {Salman}},
  \bibinfo {author} {\bibfnamefont {T.}~\bibnamefont {Neupert}}, \bibinfo
  {author} {\bibfnamefont {E.}~\bibnamefont {Morenzoni}}, \bibinfo {author}
  {\bibfnamefont {E.}~\bibnamefont {Pomjakushina}}, \bibinfo {author}
  {\bibfnamefont {F.}~\bibnamefont {von Rohr}}, \bibinfo {author}
  {\bibfnamefont {K.}~\bibnamefont {Conder}}, \bibinfo {author} {\bibfnamefont
  {G.}~\bibnamefont {Balakrishnan}}, \bibinfo {author} {\bibfnamefont {M.~C.}\
  \bibnamefont {Hatnean}}, \bibinfo {author} {\bibfnamefont {M.~R.}\
  \bibnamefont {Lees}}, \bibinfo {author} {\bibfnamefont {D.~M.}\ \bibnamefont
  {Paul}}, \bibinfo {author} {\bibfnamefont {A.}~\bibnamefont {Schilling}},
  \bibinfo {author} {\bibfnamefont {C.}~\bibnamefont {Baines}}, \bibinfo
  {author} {\bibfnamefont {H.}~\bibnamefont {Luetkens}}, \bibinfo {author}
  {\bibfnamefont {R.}~\bibnamefont {Khasanov}}, \ and\ \bibinfo {author}
  {\bibfnamefont {A.}~\bibnamefont {Amato}},\ }\href {\doibase
  10.1103/PhysRevB.89.161107} {\bibfield  {journal} {\bibinfo  {journal} {Phys.
  Rev. B}\ }\textbf {\bibinfo {volume} {89}},\ \bibinfo {pages} {161107}
  (\bibinfo {year} {2014})}\BibitemShut {NoStop}%
\bibitem [{\citenamefont {Barla}\ \emph {et~al.}(2005)\citenamefont {Barla},
  \citenamefont {Derr}, \citenamefont {Sanchez}, \citenamefont {Salce},
  \citenamefont {Lapertot}, \citenamefont {Doyle}, \citenamefont {R\"uffer},
  \citenamefont {Lengsdorf}, \citenamefont {Abd-Elmeguid},\ and\ \citenamefont
  {Flouquet}}]{Barla2005prl}%
  \BibitemOpen
  \bibfield  {author} {\bibinfo {author} {\bibfnamefont {A.}~\bibnamefont
  {Barla}}, \bibinfo {author} {\bibfnamefont {J.}~\bibnamefont {Derr}},
  \bibinfo {author} {\bibfnamefont {J.~P.}\ \bibnamefont {Sanchez}}, \bibinfo
  {author} {\bibfnamefont {B.}~\bibnamefont {Salce}}, \bibinfo {author}
  {\bibfnamefont {G.}~\bibnamefont {Lapertot}}, \bibinfo {author}
  {\bibfnamefont {B.~P.}\ \bibnamefont {Doyle}}, \bibinfo {author}
  {\bibfnamefont {R.}~\bibnamefont {R\"uffer}}, \bibinfo {author}
  {\bibfnamefont {R.}~\bibnamefont {Lengsdorf}}, \bibinfo {author}
  {\bibfnamefont {M.~M.}\ \bibnamefont {Abd-Elmeguid}}, \ and\ \bibinfo
  {author} {\bibfnamefont {J.}~\bibnamefont {Flouquet}},\ }\href {\doibase
  10.1103/PhysRevLett.94.166401} {\bibfield  {journal} {\bibinfo  {journal}
  {Phys. Rev. Lett.}\ }\textbf {\bibinfo {volume} {94}},\ \bibinfo {pages}
  {166401} (\bibinfo {year} {2005})}\BibitemShut {NoStop}%
\bibitem [{\citenamefont {Nakajima}\ \emph {et~al.}(2016)\citenamefont
  {Nakajima}, \citenamefont {Syers}, \citenamefont {Wang}, \citenamefont
  {Wang},\ and\ \citenamefont {Paglione}}]{Nakajima2016np}%
  \BibitemOpen
  \bibfield  {author} {\bibinfo {author} {\bibfnamefont {Y.}~\bibnamefont
  {Nakajima}}, \bibinfo {author} {\bibfnamefont {P.}~\bibnamefont {Syers}},
  \bibinfo {author} {\bibfnamefont {X.}~\bibnamefont {Wang}}, \bibinfo {author}
  {\bibfnamefont {R.}~\bibnamefont {Wang}}, \ and\ \bibinfo {author}
  {\bibfnamefont {J.}~\bibnamefont {Paglione}},\ }\href {\doibase
  10.1038/nphys3555} {\bibfield  {journal} {\bibinfo  {journal} {Nat. Phys.}\
  }\textbf {\bibinfo {volume} {12}},\ \bibinfo {pages} {213} (\bibinfo {year}
  {2016})}\BibitemShut {NoStop}%
\bibitem [{\citenamefont {Wolgast}\ \emph {et~al.}(2015)\citenamefont
  {Wolgast}, \citenamefont {Eo}, \citenamefont {\"Ozt\"urk}, \citenamefont
  {Li}, \citenamefont {Xiang}, \citenamefont {Tinsman}, \citenamefont {Asaba},
  \citenamefont {Lawson}, \citenamefont {Yu}, \citenamefont {Allen},
  \citenamefont {Sun}, \citenamefont {Li}, \citenamefont {Kurdak},
  \citenamefont {Kim},\ and\ \citenamefont {Fisk}}]{Wolgast2015prb}%
  \BibitemOpen
  \bibfield  {author} {\bibinfo {author} {\bibfnamefont {S.}~\bibnamefont
  {Wolgast}}, \bibinfo {author} {\bibfnamefont {Y.~S.}\ \bibnamefont {Eo}},
  \bibinfo {author} {\bibfnamefont {T.}~\bibnamefont {\"Ozt\"urk}}, \bibinfo
  {author} {\bibfnamefont {G.}~\bibnamefont {Li}}, \bibinfo {author}
  {\bibfnamefont {Z.}~\bibnamefont {Xiang}}, \bibinfo {author} {\bibfnamefont
  {C.}~\bibnamefont {Tinsman}}, \bibinfo {author} {\bibfnamefont
  {T.}~\bibnamefont {Asaba}}, \bibinfo {author} {\bibfnamefont
  {B.}~\bibnamefont {Lawson}}, \bibinfo {author} {\bibfnamefont
  {F.}~\bibnamefont {Yu}}, \bibinfo {author} {\bibfnamefont {J.~W.}\
  \bibnamefont {Allen}}, \bibinfo {author} {\bibfnamefont {K.}~\bibnamefont
  {Sun}}, \bibinfo {author} {\bibfnamefont {L.}~\bibnamefont {Li}}, \bibinfo
  {author} {\bibfnamefont {{\c{C}}.}~\bibnamefont {Kurdak}}, \bibinfo {author}
  {\bibfnamefont {D.-J.}\ \bibnamefont {Kim}}, \ and\ \bibinfo {author}
  {\bibfnamefont {Z.}~\bibnamefont {Fisk}},\ }\href {\doibase
  10.1103/PhysRevB.92.115110} {\bibfield  {journal} {\bibinfo  {journal} {Phys.
  Rev. B}\ }\textbf {\bibinfo {volume} {92}},\ \bibinfo {pages} {115110}
  (\bibinfo {year} {2015})}\BibitemShut {NoStop}%
\bibitem [{\citenamefont {Phelan}\ \emph {et~al.}(2014)\citenamefont {Phelan},
  \citenamefont {Koohpayeh}, \citenamefont {Cottingham}, \citenamefont
  {Freeland}, \citenamefont {Leiner}, \citenamefont {Broholm},\ and\
  \citenamefont {{McQueen}}}]{Phelan2014prx}%
  \BibitemOpen
  \bibfield  {author} {\bibinfo {author} {\bibfnamefont {W.~A.}~\bibnamefont
  {Phelan}}, \bibinfo {author} {\bibfnamefont {S.~M.}~\bibnamefont {Koohpayeh}},
  \bibinfo {author} {\bibfnamefont {P.}~\bibnamefont {Cottingham}}, \bibinfo
  {author} {\bibfnamefont {J.~W.}~\bibnamefont {Freeland}}, \bibinfo {author}
  {\bibfnamefont {J.~C.}~\bibnamefont {Leiner}}, \bibinfo {author} {\bibfnamefont
  {C.~L.}~\bibnamefont {Broholm}}, \ and\ \bibinfo {author} {\bibfnamefont
  {T.~M.}~\bibnamefont {{McQueen}}},\ }\href {\doibase 10.1103/PhysRevX.4.031012}
  {\bibfield  {journal} {\bibinfo  {journal} {Phys. Rev. X}\ }\textbf {\bibinfo
  {volume} {4}},\ \bibinfo {pages} {031012} (\bibinfo {year}
  {2014})}\BibitemShut {NoStop}%
\bibitem [{\citenamefont {Knolle}\ and\ \citenamefont
  {Cooper}()}]{Knolle2016arxiv}%
  \BibitemOpen
  \bibfield  {author} {\bibinfo {author} {\bibfnamefont {J.}~\bibnamefont
  {Knolle}}\ and\ \bibinfo {author} {\bibfnamefont {N.~R.}\ \bibnamefont
  {Cooper}},\ }\href {http://arxiv.org/abs/1608.02453} {\bibinfo  {journal}
  {{arXiv}:1608.02453}\ }\BibitemShut {NoStop}%
\bibitem [{\citenamefont {Morenzoni}\ \emph {et~al.}(1994)\citenamefont
  {Morenzoni}, \citenamefont {Kottmann}, \citenamefont {Maden}, \citenamefont
  {Matthias}, \citenamefont {Meyberg}, \citenamefont {Prokscha}, \citenamefont
  {Wutzke},\ and\ \citenamefont {Zimmermann}}]{Morenzoni1994prl}%
  \BibitemOpen
\bibfield  {journal} {  }\bibfield  {author} {\bibinfo {author} {\bibfnamefont
  {E.}~\bibnamefont {Morenzoni}}, \bibinfo {author} {\bibfnamefont
  {F.}~\bibnamefont {Kottmann}}, \bibinfo {author} {\bibfnamefont
  {D.}~\bibnamefont {Maden}}, \bibinfo {author} {\bibfnamefont
  {B.}~\bibnamefont {Matthias}}, \bibinfo {author} {\bibfnamefont
  {M.}~\bibnamefont {Meyberg}}, \bibinfo {author} {\bibfnamefont
  {T.}~\bibnamefont {Prokscha}}, \bibinfo {author} {\bibfnamefont
  {T.}~\bibnamefont {Wutzke}}, \ and\ \bibinfo {author} {\bibfnamefont
  {U.}~\bibnamefont {Zimmermann}},\ }\href {\doibase
  10.1103/PhysRevLett.72.2793} {\bibfield  {journal} {\bibinfo  {journal}
  {Phys. Rev. Lett.}\ }\textbf {\bibinfo {volume} {72}},\ \bibinfo {pages}
  {2793} (\bibinfo {year} {1994})}\BibitemShut {NoStop}%
\bibitem [{\citenamefont {Prokscha}\ \emph {et~al.}(2008)\citenamefont
  {Prokscha}, \citenamefont {Morenzoni}, \citenamefont {Deiters}, \citenamefont
  {Foroughi}, \citenamefont {George}, \citenamefont {Kobler}, \citenamefont
  {Suter},\ and\ \citenamefont {Vrankovic}}]{Prokscha2008nima}%
  \BibitemOpen
  \bibfield  {author} {\bibinfo {author} {\bibfnamefont {T.}~\bibnamefont
  {Prokscha}}, \bibinfo {author} {\bibfnamefont {E.}~\bibnamefont {Morenzoni}},
  \bibinfo {author} {\bibfnamefont {K.}~\bibnamefont {Deiters}}, \bibinfo
  {author} {\bibfnamefont {F.}~\bibnamefont {Foroughi}}, \bibinfo {author}
  {\bibfnamefont {D.}~\bibnamefont {George}}, \bibinfo {author} {\bibfnamefont
  {R.}~\bibnamefont {Kobler}}, \bibinfo {author} {\bibfnamefont
  {A.}~\bibnamefont {Suter}}, \ and\ \bibinfo {author} {\bibfnamefont
  {V.}~\bibnamefont {Vrankovic}},\ }\href {\doibase 10.1016/j.nima.2008.07.081}
  {\bibfield  {journal} {\bibinfo  {journal} {Nuc. Inst. Phys. A}\ }\textbf
  {\bibinfo {volume} {595}},\ \bibinfo {pages} {317} (\bibinfo {year}
  {2008})}\BibitemShut {NoStop}%
\bibitem [{\citenamefont {Yaouanc}\ and\ \citenamefont
  {R\'eotier}(2010)}]{Yaouanc2010}%
  \BibitemOpen
  \bibfield  {author} {\bibinfo {author} {\bibfnamefont {A.}~\bibnamefont
  {Yaouanc}}\ and\ \bibinfo {author} {\bibfnamefont {P.~D.~d.}\ \bibnamefont
  {R\'eotier}},\ }\href@noop {} {\emph {\bibinfo {title} {Muon Spin Rotation,
  Relaxation, and Resonance: Applications to Condensed Matter}}}\ (\bibinfo
  {publisher} {{OUP} Oxford},\ \bibinfo {year} {2010})\BibitemShut {NoStop}%
\bibitem [{\citenamefont {Suter}\ and\ \citenamefont
  {Wojek}(2012)}]{Suter2012pp}%
  \BibitemOpen
  \bibfield  {author} {\bibinfo {author} {\bibfnamefont {A.}~\bibnamefont
  {Suter}}\ and\ \bibinfo {author} {\bibfnamefont {B.}~\bibnamefont {Wojek}},\
  }\href {\doibase 10.1016/j.phpro.2012.04.042} {\bibfield  {journal} {\bibinfo
   {journal} {Physics Procedia}\ }\textbf {\bibinfo {volume} {30}},\ \bibinfo
  {pages} {69} (\bibinfo {year} {2012})}\BibitemShut {NoStop}%
\bibitem [{\citenamefont {{Ciomaga Hatnean}}\ \emph {et~al.}(2013)\citenamefont
  {{Ciomaga Hatnean}}, \citenamefont {Lees}, \citenamefont {{McK. Paul}},\ and\
  \citenamefont {Balakrishnan}}]{CiomagaHatnean2013sr}%
  \BibitemOpen
  \bibfield  {author} {\bibinfo {author} {\bibfnamefont {M.}~\bibnamefont
  {{Ciomaga Hatnean}}}, \bibinfo {author} {\bibfnamefont {M.~R.}\ \bibnamefont
  {Lees}}, \bibinfo {author} {\bibfnamefont {D.}~\bibnamefont {{McK. Paul}}}, \
  and\ \bibinfo {author} {\bibfnamefont {G.}~\bibnamefont {Balakrishnan}},\
  }\href {\doibase 10.1038/srep03071} {\bibfield  {journal} {\bibinfo
  {journal} {Scientific Reports}\ }\textbf {\bibinfo {volume} {3}},\ \bibinfo
  {pages} {3071} (\bibinfo {year} {2013})}\BibitemShut {NoStop}%
\bibitem [{\citenamefont {Morenzoni}\ \emph {et~al.}(2002)\citenamefont
  {Morenzoni}, \citenamefont {{Gl\"uckler}}, \citenamefont {Prokscha},
  \citenamefont {Khasanov}, \citenamefont {Luetkens}, \citenamefont {Birke},
  \citenamefont {Forgan}, \citenamefont {Niedermayer},\ and\ \citenamefont
  {Pleines}}]{Morenzoni2002nimb}%
  \BibitemOpen
  \bibfield  {author} {\bibinfo {author} {\bibfnamefont {E.}~\bibnamefont
  {Morenzoni}}, \bibinfo {author} {\bibfnamefont {H.}~\bibnamefont
  {{Gl\"uckler}}}, \bibinfo {author} {\bibfnamefont {T.}~\bibnamefont
  {Prokscha}}, \bibinfo {author} {\bibfnamefont {R.}~\bibnamefont {Khasanov}},
  \bibinfo {author} {\bibfnamefont {H.}~\bibnamefont {Luetkens}}, \bibinfo
  {author} {\bibfnamefont {M.}~\bibnamefont {Birke}}, \bibinfo {author}
  {\bibfnamefont {E.~M.}\ \bibnamefont {Forgan}}, \bibinfo {author}
  {\bibfnamefont {C.}~\bibnamefont {Niedermayer}}, \ and\ \bibinfo {author}
  {\bibfnamefont {M.}~\bibnamefont {Pleines}},\ }\href {\doibase
  10.1016/S0168-583X(01)01166-1} {\bibfield  {journal} {\bibinfo  {journal}
  {Nuclear Instruments and Methods in Physics Research Section B}\ }\textbf
  {\bibinfo {volume} {192}},\ \bibinfo {pages} {254} (\bibinfo {year}
  {2002})}\BibitemShut {NoStop}%
\bibitem [{\citenamefont {Uemura}\ \emph {et~al.}(1985)\citenamefont {Uemura},
  \citenamefont {Yamazaki}, \citenamefont {Harshman}, \citenamefont {Senba},\
  and\ \citenamefont {Ansaldo}}]{Uemura1985prb}%
  \BibitemOpen
  \bibfield  {author} {\bibinfo {author} {\bibfnamefont {Y.~J.}\ \bibnamefont
  {Uemura}}, \bibinfo {author} {\bibfnamefont {T.}~\bibnamefont {Yamazaki}},
  \bibinfo {author} {\bibfnamefont {D.~R.}\ \bibnamefont {Harshman}}, \bibinfo
  {author} {\bibfnamefont {M.}~\bibnamefont {Senba}}, \ and\ \bibinfo {author}
  {\bibfnamefont {E.~J.}\ \bibnamefont {Ansaldo}},\ }\href {\doibase
  10.1103/PhysRevB.31.546} {\bibfield  {journal} {\bibinfo  {journal} {Phys.
  Rev. B}\ }\textbf {\bibinfo {volume} {31}},\ \bibinfo {pages} {546} (\bibinfo
  {year} {1985})}\BibitemShut {NoStop}%
\bibitem [{\citenamefont {Keren}(1994)}]{Keren1994prb}%
  \BibitemOpen
  \bibfield  {author} {\bibinfo {author} {\bibfnamefont {A.}~\bibnamefont
  {Keren}},\ }\href {\doibase 10.1103/PhysRevB.50.10039} {\bibfield  {journal}
  {\bibinfo  {journal} {Phys. Rev. B}\ }\textbf {\bibinfo {volume} {50}},\
  \bibinfo {pages} {10039} (\bibinfo {year} {1994})}\BibitemShut {NoStop}%
\bibitem [{\citenamefont {Salman}\ \emph {et~al.}(2002)\citenamefont {Salman},
  \citenamefont {Keren}, \citenamefont {Mendels}, \citenamefont {Marvaud},
  \citenamefont {Scuiller}, \citenamefont {Verdaguer}, \citenamefont {Lord},\
  and\ \citenamefont {Baines}}]{Salman2002prb}%
  \BibitemOpen
  \bibfield  {author} {\bibinfo {author} {\bibfnamefont {Z.}~\bibnamefont
  {Salman}}, \bibinfo {author} {\bibfnamefont {A.}~\bibnamefont {Keren}},
  \bibinfo {author} {\bibfnamefont {P.}~\bibnamefont {Mendels}}, \bibinfo
  {author} {\bibfnamefont {V.}~\bibnamefont {Marvaud}}, \bibinfo {author}
  {\bibfnamefont {A.}~\bibnamefont {Scuiller}}, \bibinfo {author}
  {\bibfnamefont {M.}~\bibnamefont {Verdaguer}}, \bibinfo {author}
  {\bibfnamefont {J.~S.}\ \bibnamefont {Lord}}, \ and\ \bibinfo {author}
  {\bibfnamefont {C.}~\bibnamefont {Baines}},\ }\href {\doibase
  10.1103/PhysRevB.65.132403} {\bibfield  {journal} {\bibinfo  {journal} {Phys.
  Rev. B}\ }\textbf {\bibinfo {volume} {65}},\ \bibinfo {pages} {132403}
  (\bibinfo {year} {2002})}\BibitemShut {NoStop}%
\bibitem [{LCR()}]{LCRnote}%
  \BibitemOpen
  \href@noop {} {}\bibinfo {note} {The contribution from the level crossing
  seen at high temperature is subtracted before fitting the data. An additional
  field independent offset is also needed to fit the data, which we attribute
  to other possible sources of relaxation.}\BibitemShut {Stop}%
\bibitem [{\citenamefont {Gorshunov}\ \emph {et~al.}(1999)\citenamefont
  {Gorshunov}, \citenamefont {Sluchanko}, \citenamefont {Volkov}, \citenamefont
  {Dressel}, \citenamefont {Knebel}, \citenamefont {Loidl},\ and\ \citenamefont
  {Kunii}}]{Gorshunov1999prb}%
  \BibitemOpen
  \bibfield  {author} {\bibinfo {author} {\bibfnamefont {B.}~\bibnamefont
  {Gorshunov}}, \bibinfo {author} {\bibfnamefont {N.}~\bibnamefont
  {Sluchanko}}, \bibinfo {author} {\bibfnamefont {A.}~\bibnamefont {Volkov}},
  \bibinfo {author} {\bibfnamefont {M.}~\bibnamefont {Dressel}}, \bibinfo
  {author} {\bibfnamefont {G.}~\bibnamefont {Knebel}}, \bibinfo {author}
  {\bibfnamefont {A.}~\bibnamefont {Loidl}}, \ and\ \bibinfo {author}
  {\bibfnamefont {S.}~\bibnamefont {Kunii}},\ }\href {\doibase
  10.1103/PhysRevB.59.1808} {\bibfield  {journal} {\bibinfo  {journal} {Phys.
  Rev. B}\ }\textbf {\bibinfo {volume} {59}},\ \bibinfo {pages} {1808}
  (\bibinfo {year} {1999})}\BibitemShut {NoStop}%
\bibitem [{\citenamefont {Travaglini}\ and\ \citenamefont
  {Wachter}(1984)}]{Travaglini1984prb}%
  \BibitemOpen
  \bibfield  {author} {\bibinfo {author} {\bibfnamefont {G.}~\bibnamefont
  {Travaglini}}\ and\ \bibinfo {author} {\bibfnamefont {P.}~\bibnamefont
  {Wachter}},\ }\href {\doibase 10.1103/PhysRevB.29.893} {\bibfield  {journal}
  {\bibinfo  {journal} {Phys. Rev. B}\ }\textbf {\bibinfo {volume} {29}},\
  \bibinfo {pages} {893} (\bibinfo {year} {1984})}\BibitemShut {NoStop}%
\end{thebibliography}
\end{document}